%% file: short.tex
\documentclass[aps,prd,reprint,superscriptaddress,amsmath,amssymb,preprintnumbers,10pt,floatfix,nofootinbib]{revtex4-2}

\usepackage{dcolumn}
\usepackage{bm}
\pdfoutput=1
\usepackage[utf8]{inputenc}
\usepackage{graphicx,amsmath,amsfonts,amssymb,aas_macros,slashed}
\usepackage{bbold}
\usepackage{numprint}
\usepackage{enumitem}
 \usepackage{mathtools}

\usepackage{graphicx,color,amsmath}
\usepackage{hyperref}
\usepackage{booktabs}
\usepackage[dvipsnames]{xcolor}
\usepackage{caption}
\usepackage{subcaption}

\allowdisplaybreaks

\makeatletter
\renewcommand{\xRightarrow}[2][]{\ext@arrow 0359\Rightarrowfill@{#1}{#2}}
\makeatother

\newcommand{\keV}{\ensuremath{\mathrm{keV}}}

\newcommand{\op}[1]{\ensuremath \widehat{#1}}

\usepackage{pgfplots}

\usepackage{xparse}

\ExplSyntaxOn
\NewDocumentCommand{\stripfinalperiod}{m}
  {
    \tl_set:Nn \l_tmpa_tl {#1}
    \regex_replace_once:nnN { \s* \. \s* \Z } { } \l_tmpa_tl
    \tl_use:N \l_tmpa_tl
  }
\ExplSyntaxOff

\newif\ifarxiv 
 \arxivtrue   

\makeatletter
\ifarxiv

    \RenewDocumentCommand{\paragraph}{m}
    {%
      \refstepcounter{subsection}%
      \subsection*{\stripfinalperiod{#1}}%
    }
    
  \newcommand{\StartAppendix}{%
    \appendix
    \section*{Appendices}%
    \setcounter{subsection}{0}%
    \setcounter{equation}{0}%
    \renewcommand{\thesubsection}{\Alph{subsection}}%
    \renewcommand{\theequation}{\thesubsection\arabic{equation}}%
    \@addtoreset{equation}{subsection}%
  }

\else

  \newcommand{\StartAppendix}{%
    \setcounter{equation}{0}%
    \renewcommand{\theequation}{E\arabic{equation}}%
    \section*{End Matter}%
  }

\fi
\makeatother

\usepackage{bm}
\usepackage{mathtools}

\let\oldhat\hat

\renewcommand{\vec}[1]{\bm{#1}}
\renewcommand{\hat}[1]{\oldhat{\bm{#1}}}

\usetikzlibrary{positioning,arrows,patterns}
\usetikzlibrary{decorations.markings}
\usetikzlibrary{calc}

\tikzset{
    photon/.style={decorate, decoration={snake}, draw=black},
    vector/.style={decorate, decoration={snake}, draw},
	provector/.style={decorate, decoration={snake,amplitude=2.5pt}, draw},
	antivector/.style={decorate, decoration={snake,amplitude=-2.5pt}, draw},
    fermion/.style={draw=black, postaction={decorate},
        decoration={markings,mark=at position .55 with {\arrow[draw=black]{>}}}},
    fermionbar/.style={draw=black, postaction={decorate},
        decoration={markings,mark=at position .55 with {\arrow[draw=black]{<}}}},
    fermionnoarrow/.style={draw=black},
    gluon/.style={decorate, draw=black,
        decoration={coil,amplitude=4pt, segment length=5pt}},
    scalar/.style={dashed,draw=black, postaction={decorate},
        decoration={markings,mark=at position .55 with {\arrow[draw=black]{>}}}},
    scalarbar/.style={dashed,draw=black, postaction={decorate},
        decoration={markings,mark=at position .55 with {\arrow[draw=black]{<}}}},
    scalartwo/.style={dotted,draw=black, postaction={decorate},
        decoration={markings,mark=at position .55 with {\arrow[draw=black]{>}}}},
    scalartwobar/.style={dotted,draw=black, postaction={decorate},
        decoration={markings,mark=at position .55 with {\arrow[draw=black]{<}}}},
    scalarnoarrow/.style={dashed,draw=black},
    electron/.style={draw=black, postaction={decorate},
        decoration={markings,mark=at position .55 with {\arrow[draw=black]{>}}}},
	bigvector/.style={decorate, decoration={snake,amplitude=4pt}, draw},
    vertex/.style={draw,shape=circle,fill=black,minimum size=1pt,inner sep=0pt},
    fermion2/.style={double, draw=black, postaction={decorate},
		decoration={markings,mark=at position .55 with {\arrow[draw=black]{>}}}},
    momentum/.style={draw=black,line width=0.15mm, postaction={decorate},
        decoration={markings,mark=at position 1 with {\arrow[draw=black]{>}}}}
}

\newif\ifshowcomments
\showcommentstrue %

\ifshowcomments
  \newcommand{\jp}[1]{\textcolor{red}{JP: #1}}
  \newcommand{\ms}[1]{\textcolor{ForestGreen}{MS: #1}}
  \newcommand{\sn}[1]{\textcolor{brown}{SN: #1}}
\else
  \newcommand{\jp}[1]{}
  \newcommand{\ms}[1]{}
  \newcommand{\sn}[1]{}
\fi

\graphicspath{{figs/}}

\begin{document}


\title{Adiabatic response in the Migdal Effect}

\author{Stefan Nellen Mondragón}
\email{stefan.nellen@oeaw.ac.at}
\affiliation{Marietta Blau Institute for Particle Physics, Austrian Academy of Sciences, Dominikanerbastei 16, A-1010 Vienna, Austria}
\affiliation{University of Vienna, Faculty of Physics, Boltzmanngasse 5, A-1090 Vienna, Austria}

\author{Josef Pradler}
\email{josef.pradler@oeaw.ac.at}
\affiliation{Marietta Blau Institute for Particle Physics, Austrian Academy of Sciences, Dominikanerbastei 16, A-1010 Vienna, Austria}
\affiliation{University of Vienna, Faculty of Physics, Boltzmanngasse 5, A-1090 Vienna, Austria}

\author{Mukul Sholapurkar}
\email{mukul.sholapurkar@oeaw.ac.at}
\affiliation{Marietta Blau Institute for Particle Physics, Austrian Academy of Sciences, Dominikanerbastei 16, A-1010 Vienna, Austria}

\preprint{UWThPh  2026-5}

\begin{abstract} %
The Migdal effect---the prompt ionization induced by a sudden nuclear recoil---is widely used in direct dark matter searches, yet its validity beyond the impulse approximation has remained unresolved. We present the first first-principles calculation for isolated atoms, establishing the adiabatic crossover where ionization is suppressed. We show that this behavior is fully encoded in the scattering amplitude, without ad hoc assumptions, and map the relevant parameter space, finding that dark matter searches lie in the unsuppressed regime.
\end{abstract}

\maketitle


\paragraph{Introduction.}
The Migdal effect, in which a recoiling nucleus is accompanied by the ionization or excitation of an electron~\cite{Migdal:1941}, has emerged as a key mechanism for extending the reach of conventional nuclear recoil detectors to sub-GeV dark matter (DM) candidates interacting with nucleons~\cite{Bernabei:2007jz,Ibe:2017yqa}. This possibility is especially striking because it opens the prospect that ton-scale detectors designed for WIMP searches  become sensitive probes of light DM, hence competing with dedicated low-threshold experiments~\cite{Akerib:2018hck,XENON:2019zpr,Agnes_2023,Aalbers_2023,Huang_2023,Abe_2023}. This promise has motivated substantial theoretical effort to estimate Migdal scattering rates induced by dark matter, neutrons or neutrinos in materials with isolated free atoms, in molecules, and in semiconductors and crystals~\cite{Dolan:2017xbu,Bell:2019egg,Baxter:2019pnz,Essig:2019xkx,Liang:2019nnx,Dey:2020sai,Liu:2020pat,Knapen:2020aky,Liang:2020ryg,Flambaum:2020xxo,Bell:2021zkr,Acevedo:2021kly,Wang:2021oha,Bell:2021ihi,Cox:2022ekg,Blanco:2022pkt,Tomar:2022ofh,Berghaus:2022pbu,Li:2022acp,Bell:2023uvf,Qiao:2023pbw,AtzoriCorona:2023ais,Gu:2023pfg,Li:2023xkf,Kang:2024kec,Nakano:2024oon,Esposito:2025iry,Berghaus:2026kmj}. Several experimental proposals and attempts have also been made to directly observe the Migdal effect by scattering nuclei by neutrons~\cite{Nakamura:2020kex,Bell:2021ihi,MIGDAL:2022yip,Adams:2022zvg,Xu:2023wev,Kahn:2024nyv}, with a positive observation reported recently in a gaseous target~\cite{Yi:2026fmf}.

Yet this entire program hinges on the prompt ionization efficiency that follows the nuclear recoil, and its theoretical status is less secure than is often assumed. In  fact, existing calculations are to-date entirely based on the \textit{impulse approximation}, in which the nuclear recoil is described by a sudden momentum transfer $\boldsymbol{q}$, occurring on a timescale much shorter than any other relevant characteristic electron timescale. In the isolated atom picture, the Migdal ionization amplitude $\mathcal{M}_{fi} $ is then obtained by projecting the boosted electron cloud onto the final state, and is proportional to
\begin{align}
\label{eq:introMigdalNaive}
\widetilde{V}_{\chi N} (\vec{q}) \langle \phi_f | e^{i \frac{m_e}{M_A}\vec{q}\cdot \sum_j \op{\vec{r}}_j } | \phi_i \rangle
\simeq \frac{i}{e}\frac{m_e}{M_A} \widetilde{V}_{\chi N} (\vec{q})\ \vec{q}\cdot \vec{d}_{fi}\
\end{align}
where $ \widetilde{V}_{\chi N} (\vec{q})$ is the spatial Fourier transform of the DM-nuclear scattering potential $V_{\chi N}$ and $\vec{d}_{fi} \equiv \langle \phi_f |\sum_{j}e \op{\vec{r}}_j|\phi_i\rangle$ is the transition dipole moment between an electronic initial state $|\phi_i\rangle$ and an electronic final state $| \phi_f \rangle$.  The small ratio of electron over the atomic mass, $m_e/M_A$, reflects that the electronic excitation is induced only indirectly through the nuclear recoil. 

\begin{figure}
    \includegraphics[width=0.5\columnwidth]{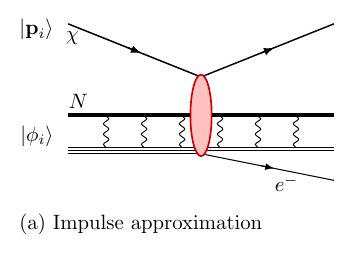}%
     \includegraphics[width=0.5\columnwidth]{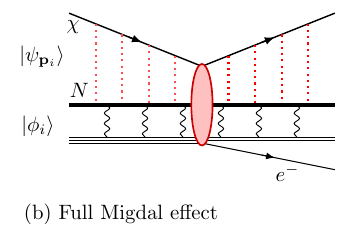}%
    \caption{(\emph{a}) $\chi N$ relative motion described by Born plane wave states $| \vec p_{i,f} \rangle$ (\emph{b})  exact  $\chi N$ states $|\psi_{\vec p_{i,f}}\rangle$ encoding the adiabatic crossover. The blob is the ionizing force operator $\frac{m_e}{M_A} (\op{\vec  r}_j \cdot \nabla \op V_{\chi N}) $ acting on the $j$th electron.}
    \label{fig:scheme}
\end{figure}

On general grounds, one expects a regime in which the momentum transfer is sufficiently slow that the electron cloud remains bound to the recoiling nucleus. In this limit, the ionization probability must be parametrically suppressed and ultimately vanish, as dictated by the adiabatic theorem. Despite the tremendous interest in the Migdal effect, however, this expectation remains largely heuristic: a derivation of the suppression, and of the breakdown of the impulse approximation, is still lacking.

In this letter, we present an ab initio calculation of the Migdal scattering rate for isolated atoms beyond the impulse approximation. This identifies the transition between the impulse regime and the adiabatic cutoff regime, where ionization ceases and the corresponding experimental sensitivity is lost. One might attempt to infer the adiabatic crossover from time-dependent scattering kernels, e.g.~by associating slow momentum transfer to the nucleus with a nontrivial temporal structure. Although such formulations may offer useful intuition, they rely on \emph{ad hoc} assignments  of timescales rather than providing a first-principles derivation. We therefore treat the system as closed, within a Hamiltonian formulation that does not require integrating out any degrees of freedom.
In this formulation, the time integration yields only the energy-conserving delta function---which has been dropped from~\eqref{eq:introMigdalNaive}---and, as we shall see, the full information about the adiabatic crossover is already encoded in the DM-nuclear scattering matrix element.


\paragraph{Formulation.}

Throughout this work we consider non-relativistic  scattering of a particle~$\chi$ with the nucleus. This is the most relevant regime to identify the adiabatic cross-over; it includes the scattering of DM and neutrons on nuclei. We also treat the nucleus as point-like, which restricts the momentum transfer of $\chi$ to tens of~MeV. This condition is amply satisfied in the phenomenologically most important Migdal regime for sub-GeV DM. (We employ natural units $\hbar = c =1$.)

It proves useful to choose so-called Jacobi coordinates of relative position with electron-nucleus relative coordinate $\vec r_j \equiv \vec x_{e,j} - \vec x_N$,  $\chi$-atom relative coordinate $\vec \rho \equiv \vec x_\chi - \vec R_{eN}$ where $\vec R_{eN}$ is the center-of-mass coordinate of the nucleus-electrons system and total center-of-mass coordinate $\vec R$ of the entire system; $\op{\vec x}_N$, $\op{\vec x}_{\chi}$, and $\op{\vec x}_{e,j}$ are the position  operators for the nucleus, $\chi$, and the $j^{\text{th}}$ electron, respectively, with $j=1,\dots,Z$ and $Z$ the atomic number. Writing the $\chi N$ interaction potential $V_{\chi N}(\op{\vec{x}}_\chi - \op{\vec{x}}_N)$ in Jacobi coordinates and expanding  to leading order in $\beta \equiv m_e/M_A < 10^{-3}$ yields
\begin{subequations}
\begin{align}
V_{\chi N}(\op{\vec{x}}_\chi - \op{\vec{x}}_N) &= V_{\chi N}\Big(\op{\vec{\rho}} + \beta \sum_{j}  \op{\vec{r}}_j\Big) \\& \simeq V_{\chi N}(\op{\vec{\rho}}) + \beta \sum_{j} \nabla_{\vec{\rho}} V_{\chi N} (\op{\vec{\rho}}) \cdot \op{\vec{r}}_j .
\label{eq:pot_expansion}
\end{align}
\end{subequations}
Importantly, it is the second term that  induces the Migdal effect by coupling electron coordinates $\vec r_j$ with $\chi$-nucleus relative coordinate~$\vec \rho$. 
Equation~\eqref{eq:pot_expansion} is equivalent to the dipole approximation made in~\eqref{eq:introMigdalNaive}; the validity of the latter has been studied in~\cite{Cox:2022ekg}.

Dropping the overall free center-of-mass motion 
the remaining total Hamiltonian reads
\begin{align}\label{eq:H_eff}
\op{H} = \op{H}_{0} + \beta \nabla_{\!\!\vec{\rho}} V_{\chi N} (\op{\vec{\rho}}) \cdot \sum_{j} \op{\vec{r}}_j,
\end{align}
where $V_{\chi N}(\op{\vec{\rho}})$, the first term of~\eqref{eq:pot_expansion}, is being absorbed into the \emph{diagonal} evolution of the scattering system, 
$\op{H}_{0} = \op{H}_{\chi N}(\op{\vec{\rho}}) + \op{H}_{\rm el}(\{\op{\vec r}_j \})$ with $\op H_{\chi N} \equiv {\op{\vec{p}}_{\rho}^2}/{(2\mu_{\chi A})} + V_{\chi N}(\op{\vec{\rho}})$. Here, $\op{\vec{p}}_{\rho}$ is the three-momentum operator canonical to $\op{\vec{\rho}}$;  $\mu_{\chi A} $ and $\mu_{eN}$ are the reduced masses of the $\chi$-atom and electron-$N$ system, respectively. The usual atomic Hamiltonian  is denoted by $\op{H}_{\rm el}(\{ \vec r_j \}) $ and it only depends on coordinates $\{ \vec r_j \}_{j=1}^Z$.

The eigenstates of $\op{H}_{0}$ form a complete basis for evaluating transition matrix elements induced by the interaction term in~\eqref{eq:H_eff}. We write them as $|\psi_{\vec{p}_{i,f}}^{\pm}\rangle \otimes |\phi_a \rangle$. The states $|\psi_{\vec{p}_{i,f}}^{\pm}\rangle$ are the stationary scattering eigenstates  $ \op H_{\chi N} |\psi_{\vec{p}_{i,f}}^{\pm}\rangle = E_{ p_{i,f}} |\psi_{\vec{p}_{i,f}}^{\pm}\rangle $ with $E_{p_{i,f}} ={p^2_{i,f}}/{2\mu_{\chi A}}$, where $\vec p_{i}$ and $\vec p_{f}$ are the free initial and final asymptotic relative momenta of $\chi$ and the atom, respectively, and $p_i$ and $p_f$ their corresponding magnitudes.
The states  $|\psi_{\vec{p}}^{\pm}\rangle$  obey incoming $(+)$ and outgoing $(-)$ scattering boundary conditions, thereby satisfying the Lippmann-Schwinger equations
$|\psi_{\vec p}^{\pm}\rangle =|\vec p\rangle+(E_{p}-\op H_{\chi N}\pm i0)^{-1}\op V_{\chi N} |\psi_{\vec p}^{\pm}\rangle$. The states $|\phi_a \rangle$ diagonalize the electronic part,  $\op{H}_{\text{el}} |\phi_a \rangle = E_{|\phi_a \rangle}  |\phi_a \rangle $ and we take $|\phi_i \rangle$ to describe a neutral atom and $|\phi_f \rangle$ a singly ionized atom plus a continuum electron of momentum~$\vec k$, conjugate to the ionized electron coordinate~$\vec{r}$. 

The Migdal ionization amplitude $\mathcal{M}_{fi}$ between initial state $|i\rangle \equiv |\psi_{\vec{p}_i}^{+}\rangle \otimes |\phi_i \rangle$ and   final state $|f\rangle \equiv |\psi_{\vec{p}_f}^{-}\rangle \otimes |\phi_f \rangle$ is given by
\begin{align}\label{eq:mat_element_1}
\mathcal{M}_{fi} &= \langle f| \beta \nabla_{\!\!\vec{\rho}} V_{\chi N} (\op{\vec{\rho}}) \cdot \sum_{j} \op{\vec{r}}_j |i\rangle 
= 
- \frac{\beta}{e} \widetilde{\vec F}_{fi}(\omega) \cdot \vec{d}_{fi}
\end{align}
where $    \widetilde{\vec F}_{fi}(\omega,q,p_i) = - \langle \psi_{\vec{p}_f}^{-} | \nabla_{\!\!\vec{\rho}} V_{\chi N} (\op{\vec{\rho}}) | \psi_{\vec{p}_i}^{+}\rangle $
is the \emph{exact} force matrix element to all orders in the $V_{\chi N}$ interaction. In the scattering process, the momentum transfer to the atom is given by $\vec{q} \equiv \vec{p}_i - \vec{p}_f$ and the electronic energy transfer is denoted by $\omega \equiv E_{|\phi_f \rangle}-E_{|\phi_i \rangle}$, which for ionization from a bound state of energy $E_{b}<0$ becomes $\omega =  \vec k^2/(2\mu_{eN}) + |E_{b}|$. Energy conservation fixes $\omega = {(p_i^2 - p_f^2)}/(2\mu_{\chi A})$. 
For centrally symmetric potentials, $|\widetilde{\vec F}_{fi}|$ is therefore a function of $\omega$, $q = |\vec q|$.
For $m_\chi\ll m_N$, the Jacobi coordinates collapse to the usual lab-frame description: the atom is initially at rest, $\chi$ has initial and final momenta $\vec p_i$ and $\vec p_f$, and the ionized electron has lab-frame momentum~$\vec k$.


\paragraph{Impulse approximation = Born limit.}

It is instructive to evaluate $ \widetilde{\vec F}_{fi} $ in the Born limit of $\chi N$ scattering.
This corresponds to replacing scattering states by plane waves:
$| \psi_{\vec{p}_i}^{+}\rangle \to |\vec{p}_i\rangle$, $| \psi_{\vec{p}_f}^{-}\rangle \to |\vec{p}_f\rangle$.  Using the commutation relation $\nabla V_{\chi N}(\widehat{\vec{\rho}}) = i [\widehat{\vec{p}}_{\rho},V_{\chi N}(\widehat{\vec{\rho}})]$, we obtain 
\begin{align}\label{eq:Born_limit}
\widetilde{\vec F}_{fi} |_{\text{Born}} = i \vec{q}\,\widetilde{V}_{\chi N} (\vec{q}) . 
\end{align}
We hence recover the previously known standard result~\cite{Ibe:2017yqa} for the Migdal amplitude in~\eqref{eq:introMigdalNaive};
note that kinematic information on $\omega$ and $p_i$ is lost in the Born limit~\eqref{eq:Born_limit}.

We thus make the important observation that the \emph{impulse approximation is automatically enforced in the Born limit}. This also matches the intuition of the Born approximation, where the interaction operator is inserted just once between the initial and final states, making the $\chi$-nucleus interaction effectively instantaneous. The leading order in the Born limit thus neglects any timescale effects on the scattering. By computing Eq.~\eqref{eq:mat_element_1} without making any additional approximations, we remove the restriction of this limit. 


\paragraph{Results for Yukawa interactions.}

\begin{figure}
\includegraphics[width =\columnwidth]{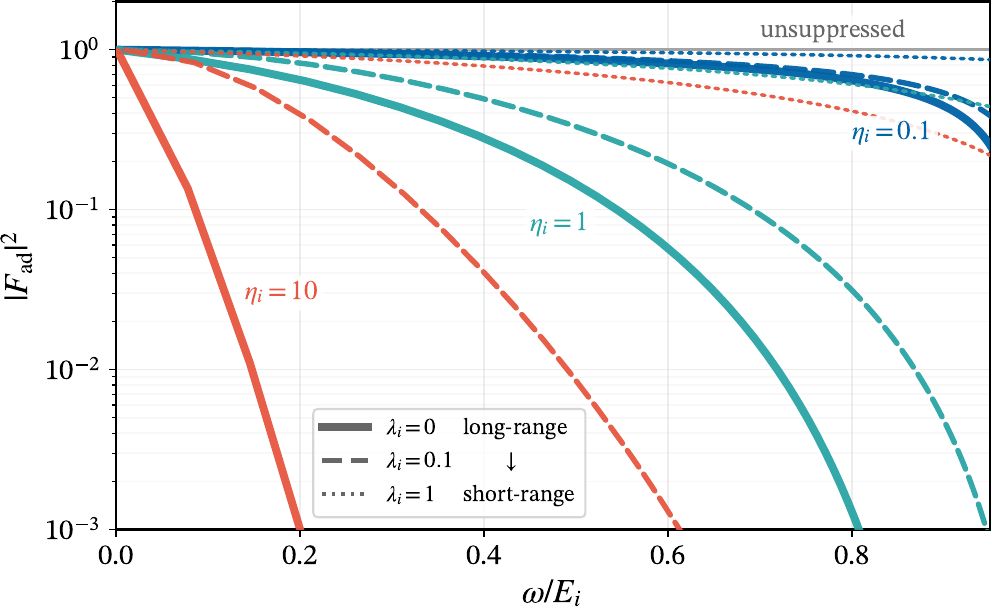}
   \caption{Squared adiabatic suppression factor $|F_{\rm ad}|^2$ as a function of fractional electronic energy transfer~$\omega/E_i$. }
    \label{fig:f_dyn}
\end{figure}

We are evidently in need of evaluating $\widetilde{\vec F}_{fi}$ to high precision, beyond the Born approximation. For this purpose, we make the most illustrative choice for $V_{\chi N}$ by that of a Yukawa potential,
$ V_{\chi N} (\rho) = ({\alpha'}/{\rho}) e^{-m_{\phi}\rho}$,
where $\alpha'$ parametrizes the strength of the interaction, and $m_{\phi}$ is the mass of the mediator coupling $\chi$ to the nucleus.
For simplicity, we consider repulsive Yukawa interactions, $\alpha' >0$.
Note that for spin-independent $\chi N$ scattering, $\alpha'$ contains a coherence factor, e.g.,~$A^2$ for isospin symmetric DM-nucleon interactions.

We define the following dimensionless parameters that characterize the dynamics of the system:
$ \eta_i \equiv {\alpha'}/{v_i}$ and  
$\lambda_i \equiv {m_{\phi}}/{p_i},$
where $v_{i} = {p_{i}}/{\mu_{\chi A}}$  is the initial relative speed of $\chi$ with respect to the center-of-mass of the atom. The parameter $\eta_i$ captures the relative strength of the interaction with respect to the speed of the $\chi$ particle. The parameter $\lambda_i$ tells if the system is in the light mediator regime ($\lambda_i \ll 1$) or in the heavy mediator regime ($\lambda_i \gg 1$). For given values of these two parameters and $m_{\chi}$, we explicitly compute the matrix element $\widetilde{\vec F}_{fi}(\omega,q,p_i)$ as a function of the kinematic variables $\omega$, $q$, and $p_i$; for details see the Appendices.

In the following, it proves useful to anchor the discussion on the limit of vanishing electronic energy transfer
\begin{align}
    \widetilde{\vec F}_{fi}(\omega = 0,q,p_i) = i \vec{q} \, \mathcal{A}_{\text{el}}(q,p_i) .
\end{align}
Here, $\mathcal{A}_\text{el}(q,p_i)$ is the \emph{exact} elastic $\chi N$ scattering amplitude. 
In the Born limit, $\mathcal{A}_\text{el}(q,p_i)$ reduces to $\widetilde{V}_{\chi N} (q)$  as can be seen from~\eqref{eq:Born_limit}. We then parameterize the magnitude of the inelastic matrix element for finite electron energy transfer~$\omega>0$ as 
\begin{align}\label{eq:fad_def}
 |\widetilde{\vec F}_{fi}(\omega,q,p_i) | = F_{\text{ad}}(\omega/E_i,q/p_i) \, q \, |\mathcal{A}_{\text{el}}(q,p_i)|,
\end{align}
with initial energy $E_i = p_i^2/(2\mu_{\chi A})$.
As we shall see below, $F_{\text{ad}}$ deserves to be called an \emph{adiabatic suppression factor}; it evidently satisfies the boundary condition  $F_\text{ad} \rightarrow 1$ for $\omega \rightarrow 0$.

Specializing to the Yukawa potential, in Fig.~\ref{fig:f_dyn}, we show $|F_\text{ad}|^2$ as a function of $\omega/E_i$ for fixed momentum transfer $q=p_i$ and  $\lambda_i = 0$ (Coulomb limit, solid lines), $\lambda_i = 0.1$ (dashed lines), and $\lambda_i = 1$ (dotted lines). Colors discriminate the choices of $\eta_i$ as labeled. The suppression increases super-exponentially with energy transfer and is strongest when entering the classical regime~$\eta_i\geq 1$.

\begin{figure}
\includegraphics[width =\columnwidth]{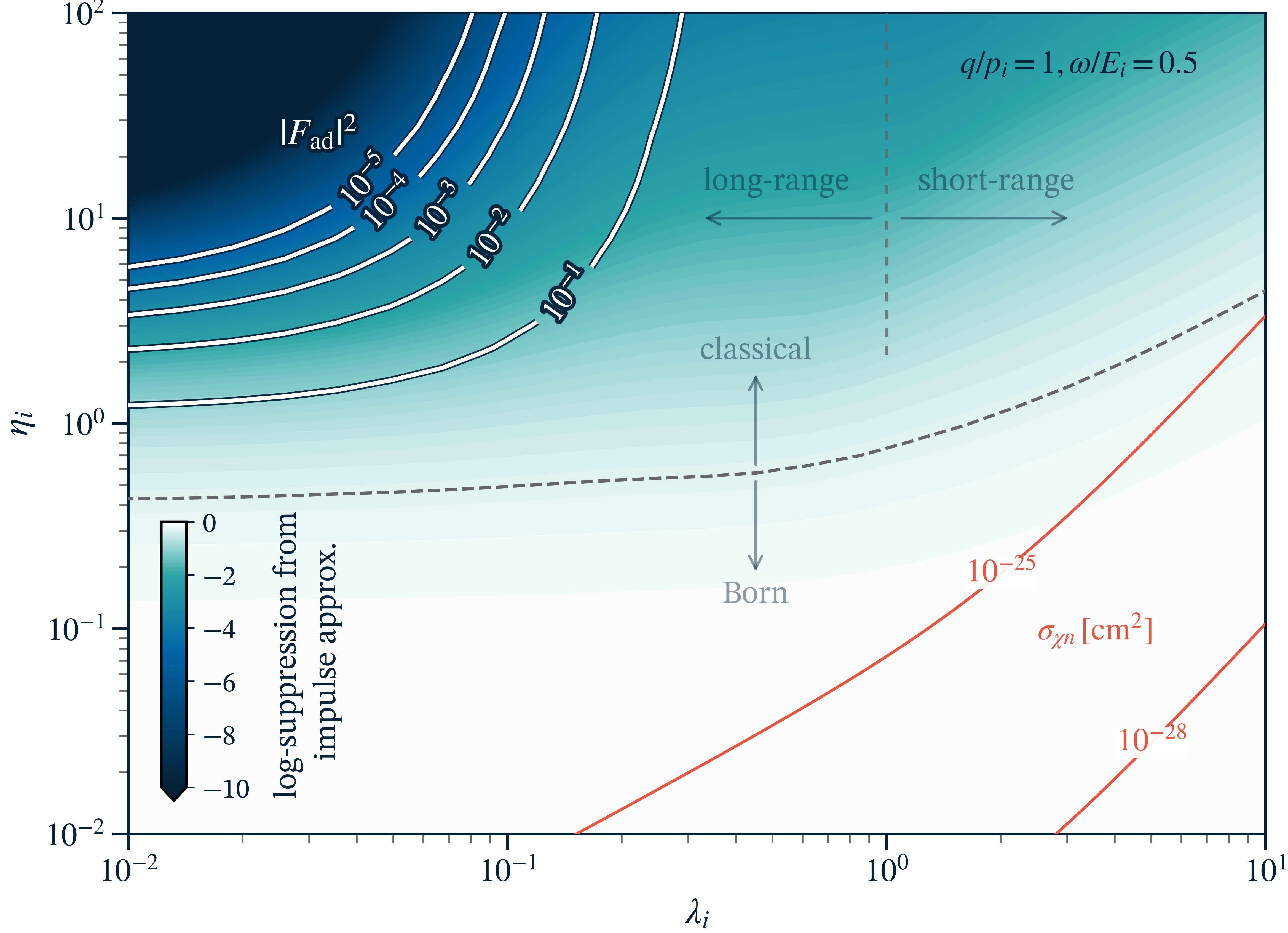}
   \caption{Log-deviation of the squared $\chi N$ scattering amplitude from the impulse approximation (Born limit) as a function of $\lambda_i$ and~$\eta_i$; the dashed contour marks a 50\% correction. The red contours show a cross section per nucleon, $\sigma_{\chi n}$, for a xenon nucleus with $p_i = 100\ \keV$.}
    \label{fig:landscape}
\end{figure}

In Fig.~\ref{fig:landscape}, we show the Yukawa parameter plane $(\lambda_i, \eta_i)$, where we indicate the various kinematical regimes by the gray dashed lines where one contour line shows the  50\% deviation from Born. The color map shows the $\log$-suppression factor of $|\widetilde{\vec F}_{fi}|^2$ relative to the squared Born amplitude/impulse approximation for $q/p_i = 1$ and $\omega/E_i=0.5$. Beyond the Born-regime, the results start  deviating significantly from the impulse approximation. This is initially due to  $|\mathcal{A}_{\text{el}}(q,p_i)| < \widetilde{V}_{\chi N} (q)$.
For long-range interactions, $\lambda_i < 1$ and in the semiclassical regime $\eta_i >1$ we then find an additional suppression given by $|F_{\rm ad}|^2$. Finally, we show by the red contours DM-nucleon cross sections $\sigma_{\chi n}$ with values as labeled for xenon in the limit $m_\chi$ being smaller than the nucleon mass for $p_i = 100~\keV$.


\paragraph{The adiabatic response regime.}

For inelastic scattering, $\omega>0$, our results show that $F_{\rm ad}$ acts as an super-exponential suppression factor.
The numerical results can be written as
\begin{align}
    |F_{\rm ad}(\omega/E_i,q/p_i)|^2
    =
    \exp\left[-\omega\,\tau_{\chi N}(\omega; q,p_i)\right] 
    \label{eq:fadexptau}
\end{align}
where we may identify $\tau_{\chi N}(\omega;q,p_i)$ as the time scale of the $\chi N$ scattering process and $\tau_{\rm ion}\equiv 1/\omega$ as the electronic transition time. Note that $\tau_{\chi N}$ has an explicit dependence on $\omega$, as the $\chi N$ scattering process involves both the initial ($E_i$) and the final ($E_f \equiv p_f^2/(2\mu_{\chi A})$) energies, and the difference between them is given by $\omega$.
When $\omega\,\tau_{\chi N}\gg 1$,  $\tau_{\rm ion}$ is short compared with the duration of the nuclear scattering process, and the corresponding Migdal transition amplitude becomes exponentially suppressed.

The adiabatic suppression is present in the large-Sommerfeld-parameter regime $\eta_i\gg 1$, implying that the classical action of the relative $\chi$-atom motion is large. Thus $\eta_i\gg 1$ is a controlled semi-classical limit in which the relative coordinate, in the elastic limit $\omega \to 0$, may be approximated by a classical trajectory, $\vec \rho=\vec \rho_{\rm cl}(t)$. 
Replacing the operator $\widehat{\vec \rho}$ in Eq.~\eqref{eq:H_eff} by $\vec \rho_{\rm cl}(t)$ produces an explicitly time-dependent electronic Hamiltonian. Although our full ab initio stationary calculation does not require this, we see now how the large-$\eta_i$ limit provides a  semi-classical interpretation of the same physics in terms of electronic response along a classical nuclear trajectory: ``accelerating the nucleus slowly'' will not induce ionization.

The emergence of a scattering time scale is impressively demonstrated in the Coulomb limit, $\lambda_i=0$. For $\eta_i\gg 1$, the de\,Broglie wavelength of the relative $\chi N$ motion, $\lambdabar= 1/p_i$, is smaller than shortest classical relative distance $\rho_0 = (2\eta_i/q) [1+q/(2p_i)]$. The maximal force experienced along the classical trajectory is then  $|\vec F_{\rm max}| = \alpha'/\rho_{0}^2$. Since the total momentum transfer $\vec{q}$ is given by $\vec{q} = \int_{-\infty}^{\infty} dt~\vec{F}(t)$, where $\vec{F}(t)$ is the time profile of the force on the classical trajectory  $\vec \rho_{\rm cl}(t)$, we parametrically expect $q = \tau_\text{cl} |\vec F_{\rm max}| $, where $\tau_\text{cl}$ is a classical scattering time scale. Note, since classical trajectories correspond to elastic scatterings, $\tau_\text{cl}$ depends only on momentum transfer $q$ and the initial momentum~$p_i$. Indeed, in the Coulomb limit we find that the full quantum-mechanical expression $\tau_{\chi N}$ in Eq.~\eqref{eq:fadexptau} asymptotes to
\begin{align}
\tau_{\chi N} (\omega;q,p_i) \to  \tau_\text{cl}(q,p_i)  \quad (\eta_i \gg 1),
\end{align}
with a constant of proportionality that is essentially unity. This proves that the suppression we find is of truly adiabatic origin.

Finally, when approaching the question of adiabaticity in the Migdal effect, the orbital time-scale of a bound electron, $\tau_{\rm orb} \sim 1/|E_b|$, is also a seemingly natural quantity to consider. This scale, however, is already encoded in the atomic ionization efficiency through the factorized dipole transition matrix element~$\vec d_{fi}$. The factorization in~\eqref{eq:pot_expansion} follows from the smallness of $\beta$ and higher-order terms that mix nuclear and electronic coordinates are parametrically suppressed by additional powers of~$\beta$. They therefore do not modify the leading adiabatic exponent. Moreover, since ionization requires $\omega \geq |E_b|$, one has $\tau_{\rm orb} \geq \tau_{\rm ion}$. Therefore, $\exp(-\tau_{\rm el}/\tau_{\rm ion}) \leq \exp(-\tau_{\rm el}/\tau_{\rm orb})$ and we conclude that~\eqref{eq:fadexptau} yields the \emph{strongest possible} suppression---and it is the one calculated from first principles.

\begin{figure}
    \centering
    \includegraphics[width = \columnwidth]{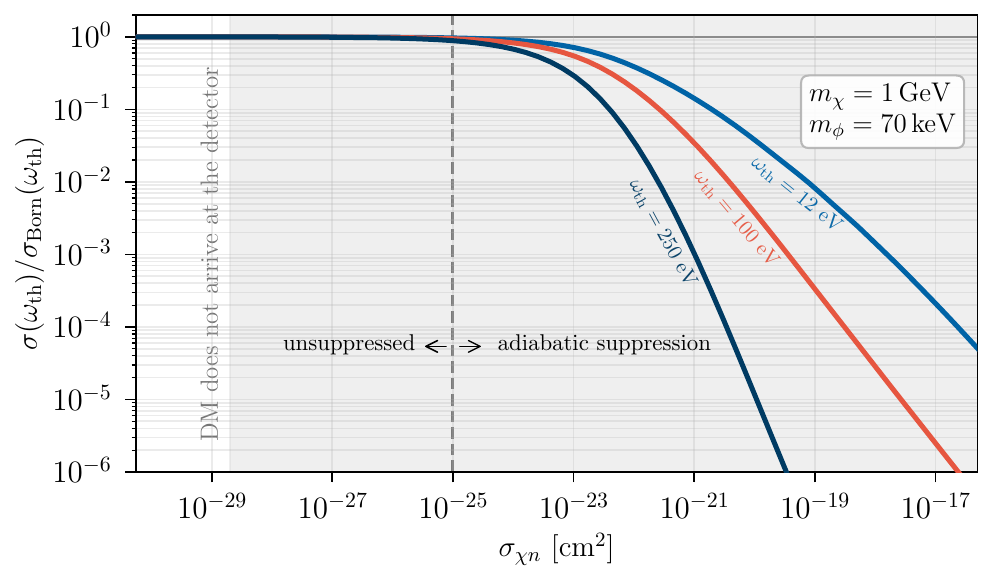}
    \caption{Integrated full Migdal ionization cross section relative to the Born-level evaluated result for $\chi$-nucleon scattering.}
    \label{fig:x-sec}
\end{figure}

\paragraph{Full ionization rates.}
Using~\eqref{eq:mat_element_1} and~\eqref{eq:fad_def} the \emph{exact} Migdal ionization rates to leading order in~$\beta$ can still be written in a factorized form, in the way they have been broadly applied in the literature. For $m_\chi \ll m_N$, the Jacobi coordinates coincide with the lab frame with nuclear recoil energy given by $E_R = q^2/(2 m_N)$ and the electronic energy $\omega$ is the sum of electron recoil energy $\vec k^2/(2m_e)$ and binding energy~$|E_{n\ell}|$.
The double differential rate is then,
\begin{equation}\label{eq:dsigdedk-lab}
       \frac{ \mathrm d \sigma_{n \ell \rightarrow k}}{\mathrm d E_R d\omega} = |F_\text{ad}|^2 | \,F_\text{el}|^2\,\frac{\mathrm d \sigma_{\chi N}}{\mathrm dE_R}\bigg|_\text{Born} \,\frac{1}{2\pi}\frac{\mathrm d p_{n \ell \rightarrow k}}{\mathrm d \omega},
\end{equation}
where $\left. \mathrm d\sigma_{\chi N}/\mathrm dE_R \right|_\text{Born}$ is the differential elastic scattering cross section in the Born approximation. The factor is ${\mathrm d p_{n \ell \rightarrow k}/\mathrm d\omega}$ the energy-differential ionization probability for an electron  with principal and orbital quantum numbers $n$ and $\ell$ obtained from~$\vec d_{fi}$, and it has computed in the literature to high precision~\cite{Ibe:2017yqa,Cox:2022ekg}.
Hence full account on corrections beyond the Born limit and from adiabatic suppression is given by the multiplicative squared form factors $| \,F_\text{el}|^2$ and $|F_\text{ad}|^2$, respectively. The former is defined through $|F_\text{el}|^2\equiv {|\mathcal{A}_{\text{el}}(q,p_i)|^2}/{|\widetilde{V}_{\chi N} (q)|^2}$; unlike $F_\text{ad}$, the elastic from factor is not a function of $\omega$. Fig.~\ref{fig:x-sec} shows the ratio of the exact total cross section, integrated from an energy threshold~$\omega_{\rm th}$, to the corresponding standard result obtained in the impulse approximation. One observes that adiabatic suppression is realized in a regime where terrestrial detectors are shielded from~DM.

\paragraph{Conclusions and outlook.}

In this work, we present a first-principles calculation of atomic Migdal scattering beyond the impulse approximation. We identify the adiabatic crossover into suppressed ionization in the semiclassical regime of long-range $\chi N$ scattering. The central phenomenological result is that existing sub-GeV direct detection searches operate in the unsuppressed regime: the limits derived using the standard Migdal treatment are therefore robust. Our analysis thus removes a major theoretical assumption behind previous Migdal searches and establishes their validity from first principles.

Our results also clarify the role of calibration and complementary probes. Neutrino- and photon-recoil induced Migdal scattering remain unsuppressed. Neutron scattering, on the other hand, can probe dynamics beyond the Born limit, making it a particularly important setting where the effects identified here should be investigated. Finally, an important next step is to translate this formulation to solid-state detectors, where Migdal processes have so far been described using rather different theoretical frameworks. These questions define natural extensions of the present work and will be addressed separately~\cite{upcoming}.


\medskip
\paragraph{Acknowledgements.}
Funded by the European Union (ERC, NLO-DM, 101044443). This work was also supported by the Research Network Quantum Aspects of Spacetime (TURIS). We acknowledge the financial support by the Vienna Doctoral School in Physics (VDSP).


\ifarxiv
  \StartAppendix
  \input{appendix.tex}

\input{short.bbl}
\else

\input{short.bbl}
  \StartAppendix

\input{appendix.tex}
\fi


\end{document}

%% file: appendix.tex

Here we provide details on the numerical evaluation of the force matrix elements, analytical results for  $\chi N$  scattering in the Coulomb limit, and the definition of the total cross section used in~Fig.~\ref{fig:x-sec}.


\subsection{Numerical Computation of Force Matrix Elements}
To effectively compute the force matrix elements $\widetilde{\vec{F}}_{fi} $ numerically, we perform a partial wave expansion of the scattering wave functions $\psi_{\vec{p}_\rho}^\pm(\vec \rho) =  \langle\vec \rho| \psi_{\vec{p}_\rho}^\pm\rangle$.
For a finite-range potential, these wave functions are given by
\begin{equation}
\label{eq:partialwave}
\psi_{\vec{p}_\rho}^\pm(\vec \rho) \simeq \sum_{\ell =0}^{\ell_{\rm max}} \sum_{m = -\ell}^\ell C_{\ell m}^{\pm}(\vec{p}_\rho)   R_{p_\rho \ell}(\rho)Y_{\ell m}(\hat{\vec{\rho}}),
\end{equation}
where $ R_{p_\rho \ell}(\rho)$ and $Y_{lm}(\hat{\vec{\rho}})$ are the radial and angular (spherical harmonic) wavefunctions, respectively; small hats---in contrast to the wide hats used for operators in the main text---denote the direction of vectors. 
The coefficients
\begin{equation}
    C_{\ell m}^{\pm} (\vec{p}_\rho)= \frac{4\pi i^{\ell} e^{\pm i\delta_\ell(p_{\rho})}~ Y_{\ell m}^{*}(\hat{\vec{p}}_\rho)}{p_{\rho}}
\end{equation}
enforce the correct incoming (outgoing) plane wave boundary conditions.
The matrix element then reads
\begin{align}\label{eq:Fxplcit}
     \widetilde{\vec{F}}_{fi}(\omega,q,p_i)&= -
     \sum_{\ell,\ell'}\sum_{m,m'}
     \frac{(2\ell'+1)(2\ell+1)i^{\ell-\ell'}}{p_ip_f} \nonumber \\ 
     & \times e^{i\left[\delta_\ell(p_i)+\delta_{\ell'}(p_f)\right]}  \, \mathcal{I}^{R}_{\ell' p_f; \ell p_i}  \, \langle \ell' m' | \op{\hat {\vec \rho}} | \ell  m \rangle.
\end{align}
This matrix element contains the scattering phase shifts $\delta_{\ell}(p_\rho)$,  the radial integral
\begin{equation}
    \mathcal{I}^{R}_{\ell' p_f;\ell p_i} = \int_0^\infty \mathrm{d} \rho \, \rho^2 \, \frac{\mathrm{d}V_{\chi N}(\rho)}{\mathrm{d}\rho} \, R_{{p_f} \ell'} ^*(\rho) \,R_{{p_i} \ell}(\rho).
\end{equation} 
and the angular transition dipole
\begin{align}
    \langle \ell' m_\ell' | \op{\hat {\vec \rho}} | \ell  m_\ell \rangle & = (-1)^{m'} \sqrt{(2 \ell +1)(2 \ell' + 1)} \nonumber \\
    & \times \begin{pmatrix}
        \ell' & \ell & 1\\
        0 &  0 & 0
    \end{pmatrix}
    \begin{pmatrix}
        \ell' & \ell & 1\\
        m' &  m'-m & m
    \end{pmatrix}.
\end{align}
Selection rules enforce $\ell' = \ell \pm 1$ and $m_\ell' = m, \, m\pm 1$.

The radial integrals are computed numerically.
To this end, we first solve the radial Schrödinger equation for the Yukawa problem in the continuum. 
The solutions are then matched asymptotically to free partial waves 
$R_{{p_{\rho}} \ell} (\rho \rightarrow \infty) \rightarrow\sin{\left[p_{\rho}\rho - \frac{\pi}{2}\ell + \delta_\ell(p_{\rho})\right]}/\rho$, which provides the phase shift and the correct Dirac-delta normalization.
A \emph{log-derivative Johnson} method~\cite{Johnson:1973ue} is used to solve the radial Schrödinger equation for each incoming/outgoing $\ell' = \ell\pm1$ pair, which is then used to perform the radial integral, and finally summed up in the matrix element, as given by \eqref{eq:Fxplcit}, until convergence is reached.
Stability and convergence of the numerical solution were tested by varying the asymptotic matching radius, the step-size and the partial wave cutoff~$\ell_\text{max}$ in the summation of~\eqref{eq:partialwave}.

The latter scales as  $\ell_\text{max}\sim 1/\lambda_i$; note that $\lambda_f > \lambda_i$ where $\lambda_f = m_{\phi}/p_f$ so that $\ell_\text{max}$ based on $\lambda_i$ covers both, incoming and outgoing waves.
In either case, the $\lambda_{i,f} =0 $ limit cannot be treated numerically.
An exact analytic solution to this limit is provided in the following subsection.


\subsection{The Coulomb Force Matrix Element}
The force matrix element $\widetilde{\vec{F}}_{fi}$ can be calculated analytically in the Coulomb limit.  
For a Coulomb potential the scattering states are known analytically, the so-called Coulomb wavefunctions
\begin{equation}
    \psi_{\vec{p}_\rho}^{(\pm)}(\vec \rho) = \Gamma(1 \pm i \eta) e^{- \frac{\pi}{2}\eta}e^{i \vec{p}_\rho \cdot \vec{\rho}} {}_1F_1(\mp i \eta, \, 1,\, i p_\rho \rho \mp i \vec p_\rho \cdot \vec{\rho}),
\end{equation}
where ${}_1F_1$ is a confluent hypergeometric function, also know as Kummer's function.
By making use of the following identities $\nabla \op{V}_{\chi N} = - i[\op{H}_{\chi N}, \, \op{\vec{p}}_\rho]$ and  $\op{\vec p}_\rho = i \mu_{\chi A} [\op H_{\chi N}, \, \op{\vec \rho}]$ we can write the force matrix element as a dipole
\begin{equation}
    \widetilde{\vec{F}}_{fi}(\omega,q,p_i) = -\frac{(p_f^2 -p_i^2)^2}{4 \mu_{\chi A}}\langle \psi^-_{\vec{p}_f} | \op{\vec{ \rho}}| \psi^+_{\vec{p}_i} \rangle.
\end{equation}
This manipulation holds for any potential, not only the Coulomb potential.

Using the Coulomb wavefunctions, the matrix element can be written as a spatial integral
\begin{align}
    \widetilde{\vec{F}}_{fi}(\omega,q,p_i) & = -\frac{(p_f^2 -p_i^2)^2}{4 \mu_{\chi A}} \Gamma(1 + i \eta_i) \Gamma(1 + i \eta_f)\nonumber \\
    & \times  e^{-\frac{\pi}{2}(\eta_i + \eta_f)} \int \mathrm{d}^3\rho \, \Big[\vec{\rho} \, e^{-i(\vec p_f - \vec{p}_i)\cdot \vec \rho} \nonumber \\
    & \times {}_1F_1(-i \eta_f, \, 1,\, i p_f \rho -i \vec p_f \cdot \vec{\rho}) \nonumber \\
    & \times {}_1F_1(- i \eta_i, \, 1,\, i p_i \rho - i \vec{p}_i \cdot \vec{\rho}) \Big].
\end{align}
We note that $p_f$ and $\eta_f$ are functions of $\omega$ and $p_i$. 
Dependence on $q$ enters through the relative direction of the two momentum vectors: $q^2 = p_i^2 + p_f^2 - 2 \vec{p}_i \cdot \vec{p}_f$.
The integration is performed using a generating function, also known as the Nordsieck integral~\cite{Nordsieck:1954zz},
\begin{equation}
    \mathcal{I}_N(\nu, \, \vec p_i, \, \vec p_f, \, \vec q) = \int \mathrm{d}^3 \rho \, e^{-\nu \rho}\frac{e^{i \vec q \cdot \vec \rho}}{\rho} {}_1^{} F_1^{(i)}(\vec{\rho}) {}_1^{} F_1^{(f)}(\vec{\rho}),
\end{equation}
 where ${}_1^{}F_1^{(i/f)}(\vec{\rho}) = {}_1F_1(-i \eta_{i/f}, \, 1,\, i p_{i/f} \rho -i \vec p_{i/f} \cdot \vec{\rho})$.
The Nordsieck integral is given in closed form as
\begin{align}
    \mathcal{I}_N(\nu, \, \vec p_i, \, \vec p_f, \, \vec q) = & \frac{2 \pi}{a} e^{-\pi\eta_i} \bigg(\frac{a}{c}\bigg)^{i \eta_i} \bigg(\frac{c + d}{c}\bigg)^{-i \eta_f} \nonumber \\
    & \times {}_2F_1\bigg(1 - i \eta_i, \,i \eta_f, \, 1, \frac{a d- bc}{a(c+d)}\bigg).
\end{align}
The different constituents of this function are given by 
\begin{align}
    a & =\frac{1}{2}(q^2 + \nu^2) , & b & = \vec{p}_f \cdot \vec q - i \nu p_f, \nonumber \\
    c & =\vec p_i \cdot \vec q + i\nu p_i - a  , & d & = p_i p_f + \vec p_i \cdot \vec p_f -b,
\end{align}
and ${}_2F_1$ denotes the standard or Gaussian hypergeometric function. The force matrix element is then obtained by taking the following limit
\begin{align}
   \widetilde{\vec{F}}_{fi}(\omega,q,p_i) =&- \frac{(p_f^2 -p_i^2)^2}{4 \mu_{\chi A}}\Gamma(1 + i \eta_i) \Gamma(1 + i \eta_f) \nonumber\\
    & \times  e^{-\frac{\pi}{2}(\eta_i + \eta_f)}\lim_{\nu\rightarrow 0^+} \lim_{\vec{q} \rightarrow\vec{p}_i - \vec{p_f}} \partial_\nu \nabla_{\!\!\vec q} \mathcal{I}_N.
\end{align}
A lengthy simplification of the limit yields the following form of the matrix element
\begin{align}\label{eq:sfi-coul}
    \widetilde{\vec{F}}_{fi}& (\omega,q,p_i) =   
    \frac{ 4 \pi e^{(\eta_i - \eta_f) \pi/2} }{\mu (z-1)(p_i - p_f)^2}  \nonumber \\
     \times & \Gamma(1 +i \eta_i) \Gamma(1+i \eta_f) \left[(1-z)\frac{p_i-p_f}{p_i+p_f}\right]^{-i( \eta_i + 
\eta_f)}   \nonumber \\ 
  \times & \Big[\eta_i p_i \,{}_2F_1(z) (\vec{p}_f - \vec{p}_i) + i(1-z) {}_2^{}F_1'(z) (p_i \vec{p}_f - p_f \vec{p}_i) \Big],
\end{align}
with $  z = {-2(p_i p_f - \vec{p}_i \cdot \vec{p}_f)}{(p_i - p_f)^{-2}}$.
Here, we use the short\-hand notation $
    {}_2F_1(z) = {}_2F_1(-i \eta_i, \, -i \eta_f, \, 1, z)$
and   ${}_2^{}F_1'(z)  \equiv \partial_z \, {}_2F_1(i \eta_i, \, i \eta_f, \, 1, z) $, which is explicitly $ {}_2^{}F_1'(z)=- \eta_i \eta_f  {}_2F_1(i \eta_i +1, \, i \eta_f+1, \, 2, z)$.


\subsection{Total cross section}

In Fig.~\ref{fig:x-sec} we plot ratios of total cross sections. They are obtained by integrating over the recoil and electron energies, and subsequently summing over the available shells for a given $\omega$-threshold
\begin{equation}
    \sigma(\omega_\text{th}) = \sum_{(n,\ell)}\int_{\max(\omega_\text{th},E_{n\ell})}^{p_i^2/2m_\chi} \int_{q_-^2/2 m_N}^{q^2_+/2 m_N}\mathrm{d}E_R \mathrm{d}\omega \, \frac{\mathrm{d}\sigma_{n \ell \rightarrow k}}{\mathrm{d}E_R \mathrm{d}\omega},
\end{equation}
where $q_\pm = p_i \pm p_f$.